\documentclass[ajp,reprint]{revtex4-1} 
\usepackage{bm}
\usepackage{graphicx}
\usepackage{dsfont}

\usepackage{bm}
\usepackage{mathrsfs}

\usepackage{framed}
\usepackage{multirow}
\usepackage[all]{xy}
\usepackage{graphicx}
\usepackage{framed}
\usepackage{comment}
\usepackage{here}                    % manages figures
\usepackage{latexsym}		     % to get LASY symbols
\usepackage{amsmath}     
\usepackage{amsfonts}  
\usepackage{amssymb}
\usepackage{amsthm}
\usepackage{dcolumn}
\usepackage{color}
\usepackage{mathrsfs}

\renewcommand{\phi}{\varphi}
\renewcommand{\>}{\rangle}
\newcommand{\<}{\langle}

\newcommand{\ket}[1]{|#1\>}
\newcommand{\bra}[1]{\<#1|}
\newcommand{\braket}[2]{\langle #1 | #2 \rangle}

\newcommand{\be}{\begin{equation}}
\newcommand{\ee}{\end{equation}}
\newcommand{\bea}{\begin{eqnarray}}
\newcommand{\eea}{\end{eqnarray}}

\newcommand{\Int}{\mathbb{Z}}

\newcommand{\olap}[2]{\langle #1 | #2 \rangle}

\renewcommand{\phi}{\varphi}

\usepackage[colorlinks,citecolor=blue,linkcolor=blue,urlcolor=blue]{hyperref}

\begin{document}

% Be sure to use the \title, \author, \affiliation, and \abstract macros
% to format your title page.  Don't use lower-level macros to  manually
% adjust the fonts and centering.

\title{Minimum-error state discrimination and Fano's inequality}
% In a long title you can use \\ to force a line break at a certain location.

%When submitting the manuscript for review, do not include the author's name or institution
%\author{Daniel V. Schroeder}
%\email{dschroeder@weber.edu} % optional
%\altaffiliation[permanent address: ]{101 Main Street, Anytown, USA} % optional second address
% If there were a second author at the same address, we would put another 
% \author{} statement here.  Don't combine multiple authors in a single
% \author statement.
%\affiliation{Department of Physics, Weber State University, Ogden, UT 84408-2508}
% Please provide a full mailing address here.

\author{Georgios M. Nikolopoulos}
%\email{nikolg@iesl.forth.gr}
\affiliation{Institute of Electronic Structure \& Laser, FORTH, GR-70013 Heraklion, Greece}
\affiliation{Center for Quantum Science \& Technologies (FORTH-QuTech), GR-70013 Heraklion, Greece}

% See the REVTeX documentation for more examples of author and affiliation lists.

\date{\today}

\begin{abstract}
The discrimination between non-orthogonal quantum states plays a pivotal role in quantum information processing and quantum technology.  
Strategies that minimize the error probability are of particular importance, but they are only known for special classes of problems. 
Certain forms of Fano's inequality yield a bound on the error probability, but it is not known how close this bound is to the minimum-error probability achieved by means of optimal measurements. 
In this work we  discuss how the minimum-error probability compares to the error bound obtained through the Fano's inequality for several scenarios, some of which are amenable to analytic treatments. 
\end{abstract}

\maketitle % title page is now complete

\section{Introduction}

In quantum information processing and quantum technology,  information is encoded in the  state of quantum systems. 
Quantum computing typically deals with large quantum systems consisting of many physical qubits, and the main aim is to solve numerical problems or simulate processes more quickly than is possible with classical computers (e.g, see chapters 5 and 6 in Ref. \cite{NCbook}).  
In quantum computing, the entire system is prepared in some initial state, and a quantum algorithm defines the type and the sequence of quantum operations (gates) that are applied on the qubits, thereby transforming the state of the quantum system.  
Any quantum computation/simulation should return an output that is associated with the final state of the quantum computer and can be retrieved only by means of quantum 
measurements . The outcomes of such measurements are classical quantities (e.g., a click in a detector), and they essentially connect the quantum world to our classical one. In the case of quantum communication and cryptography, the situation is somewhat different, but  quantum measurements play a central role in these cases as well. For instance, in a typical quantum key-distribution scenario, two users exchange quantum states chosen at random and independently from a publicly known set of non-orthogonal states (e.g, see section 12.6 in Ref. \cite{NCbook}). The transmitted quantum states are measured independently, and the outcomes are processed classically in order the two users to end up with a common secret numerical key about which a potential adversary has no information. 

We see therefore that in the field of quantum information processing and quantum technology one faces the following fundamental problem: A quantum system is prepared in a state that belongs to a set of publicly known quantum states. If the state is initially unknown, how well can one determine which state it is? When the set involves mutually orthogonal states, the task is straightforward, and one can learn exactly the state of the system by performing 
a projective measurement along the orthogonal directions related to the given set of states. The problem becomes non trivial, and far more interesting, when the possible quantum states are not mutually orthogonal. In this case, they cannot be discriminated perfectly, and information can be obtained only at the cost of permanently disturbing the state of the given quantum system. 
Thus, the problem reduces to an optimization problem with respect to some appropriately chosen figure of merit. The most natural goal is to minimize the probability for making an error in deducing the unknown state of the given system. 

The minimum-error discrimination problem is not easy, and it has been the subject of research for about 50 years. An extensive review of the current status  of the field is beyond the scope of the present work, and 
the interested reader may refer to related review articles \cite{review1,BergouReview,BaeKwekReview}, textbooks \cite{Barnettbook,BHbook}, or  theses \cite{WeirThesis}. 
The necessary and sufficient conditions that must be satisfied by a probability-operator measure (POM) for achieving minimum error in distinguishing between the states $\{ \hat{\rho}_i \}$ occurring with probabilities $\{p_i\}$ are known \cite{review1,WeirThesis}, 
but the details of the optimal measurement are known for certain special cases, where a few states are involved \cite{helstrom76,AndBarGilHun02}, or the states 
posses some sort of symmetry and/or are equiprobable \cite{review1,Ban97,Bar01,HauWoo94}. In more general forms, the problem can be solved by means of geometric \cite{geom1,geom2,geom3} and algebraic methods \cite{WeiBarCroPRA17}, or through semidefinite programming \cite{semidef}.  

Fano's inequality is a fundamental inequality of information theory, applicable to hypothesis-testing and decision-making problems \cite{CTbook}. In the framework of quantum-state discrimination, it provides a lower bound on the error probability in deducing the unknown state of the given quantum system.  \cite{NCbook,Wildebook,SWbook} 

The problem of minimum-error discrimination and Fano's inequality constitute an integral part of modern textbooks on quantum theory, quantum optics and quantum information (e.g., see \cite{NCbook,Wildebook,SWbook}), while they are also included in related 
undergraduate  and graduate courses in many universities and colleges around the world. 
In most cases, the discussion is limited to a proof of Fano's inequality, and there is no attempt to compare the predictions of the 
inequality to the minimum-error probability. Hence, students interested in these topics may wonder how the error bound obtained through Fano's inequality compares to the 
minimum-error probability, or whether there are 
settings in which Fano's inequality yields the minimum-error probability.

In the present work, these questions are addressed in scenarios that are of practical importance and are fully or partially amenable to analytic solutions. 
To the best of our knowledge, the questions under consideration have not been addressed in research papers, and thus our work may be of interest for researchers as well as educators.  

\section{Formalism} 
\label{sec2} 

Consider a system and a set of pure quantum states 
\begin{subequations}
\be
{\mathbb S}_N \equiv \{\ket{\psi_i}~:~i \in \Int_N\},
\label{set:eq}
\ee
with $\Int_N:=\{0,1, \ldots N-1\}$ and $N>1$. 
The quantum state $\ket{\psi_i}$ is fully identified by the integer $i$. 
 The set is publicly known, and  
for any distinct $i$ and $i^\prime$ in $\Int_N$ we have 
\be
|\braket{\psi_{i^\prime}}{\psi_i}| < 1.
\ee
\end{subequations}

The problem of state discrimination is as follows. We are given the quantum system prepared in one of the states $\ket{\psi_{i} }\in {\mathbb S}_N$. We also know in advance the probabilities $p_i$ that the system has been prepared in each of the states $\ket{\psi_i}$, where $\sum_i p_i = 1$. 
Our task is to infer the state of the quantum system (i.e., the label $i\in\Int_N$) by applying a measurement.

\subsection{Fano's inequality and its variants}
\label{sec2a}

Fano's inequality relies on fundamental principles of information theory, and it is applicable to multiple-hypothesis-testing and decision-making problems, including the state-discrimination problem. 
We will assume that the outcome $j\in\Int_N$ of the measurement is our guess for the label $i$ of the given state, with no  further processing needed. The random variable $j$ is related to 
the label $i$ though the conditional probability 
 $P(j|i)$ for obtaining outcome $j$ given that the system is prepared in state $\ket{\psi_i}$, where this conditional probability depends on the measurement. 
Our initial uncertainty about the label of the given state is expressed through the Shannon entropy $H(i):=-\sum_i p_i\log_2(p_i)$, whereas after the measurement the uncertainty is expressed through the conditional entropy, which is given by 
\bea
H(i|j)&:=& -\sum_{i,j} p_i P(j|i) \log_2\left [  \frac{P(j|i) p_i}{q_j} \right ]
\label{cond_ent:def2}
\eea
with 
\bea
q_j = \sum_i P(j|i) p_i.
\label{q_y:eq}
\eea
This conditional entropy is related to the initial entropy $H(i)$  through the relation 
\bea
H(i|j) = H(i) - I(i:j), 
\label{H_I:eq}
\eea
where $I(i:j)$ is the information gain. 

We wish to bound the average probability of error $p_{\rm err} : = {\rm Pr}(j\neq i)$.
In general, the conditional entropy $H(i|j)$ is zero if and only if $i$ is a function of $j$, meaning that we can obtain $i$ from $j$ with certainty if and only if $H(i|j) = 0$ or equivalently the  information gain is maximized, i.e., $I(i:j) = H(i)$. 
Hence, we expect that the probability of error will be small only if the conditional entropy $H(i|j)$ is small. 
Fano's inequality quantifies this intuition, stating  that 
\be
H_{\rm bin}(p_{\rm err}) +p_{\rm err}\log_2(N-1)\geq H(i|j),
\label{fano1:eq}
\ee
where $H_{\rm bin}(p_{\rm err}) := -p_{\rm err}\log_2(p_{\rm err})-(1-p_{\rm err})\log_2(1-p_{\rm err})$ is the binary entropy. 

The proof of Fano's inequality can be found in standard textbooks \cite{NCbook,CTbook}, and an intuitive explanation of the inequality can be given in the framework of a scenario pertaining to two parties, Alice and Bob. 
Alice chooses at random the label $i$ and prepares the state $\ket{\psi_i}$, which is given to Bob. Bob performs a measurement on the state and obtains $j$, which is also sent to Alice. 
The conditional entropy $H(i|j)$ is the average number of bits that Alice has to transmit in order for Bob to infer the label $i$ of the state from his outcome $j$. 
In particular, we can imagine Alice communicating first some bits, in order to inform Bob whether  $j=i$ or not. 
The related distribution for this is $\{p_{\rm err},1-p_{\rm err}\}$, and she has  to transmit $H_{\rm bin}(p_{\rm err}) $ bits on average \cite{footnote}.
When $i\neq j$, which  occurs with probability $p_{\rm err}$, any of the $(N-1)$ different values are possible. 
In this case Alice needs to communicate at most $\log_2(N-1)$ bits of information in order for Bob to identify $i$. 
As a result,  the conditional entropy $H(i|j)$ is expected to be bounded from above by the quantity on the left-hand-side of inequality (\ref{fano1:eq}).

It is worth recalling here that perfect discrimination of mutually orthogonal quantum states is always possible, and in this case $p_{\rm err} = 0$. 
In most problems of quantum information processing, however, one typically has to discriminate between non-orthogonal quantum states, and thus $H(i|j) \neq 0$, which in turn implies $p_{\rm err} > 0$. 
For a given measurement, one has to calculate $H(i|j) $ and solve inequality (\ref{fano1:eq}) in order to obtain a lower bound on $p_{\rm err}$. 
 
 Taking into account Eq. (\ref{H_I:eq}) and the Holevo bound \cite{holevo73}, one can obtain a rather general (yet weaker) inequality for $p_{\rm err}$, which does not require any prior knowledge of the applied measurement.
The Holevo information \cite{NCbook,holevo73} is defined by 
\bea
\chi := S\left (\hat{\rho} \right ) - \sum_{i\in \Int_N} 
p_{i} S\left ( \hat{\rho}_{i} \right ),
\label{holevoEq1a}
\eea 
where
 $S(\sigma):=-{\rm Tr}[\sigma\log_2(\sigma)]$ is the von Neumann entropy,  and 
\bea
\label{holevoEq1b}
&&\hat{\rho}_{i} := \ket{\psi_{ i} }\bra{\psi_{i} },\\
&&\hat{\rho} := \sum_{{i} \in \Int_N} p_{ i}  \hat{\rho}_{i}.
\label{holevoEq1c}
\eea
 It sets an upper bound on the information $I(i:j)$ that can be extracted from the quantum system when it is prepared in a state chosen at random from the set ${\mathbb S}_N$ defined in Eq. (\ref{set:eq}). In particular, we have  
\begin{subequations}
\be
I(i:j)\leq \chi. 
\label{holevoIneq}
\ee
\end{subequations}
For pure states, the second term in Eq. (\ref{holevoEq1a}) vanishes giving $\chi = S\left (\hat{\rho} \right ) $. 
The Holevo bound does not make any assumptions about the measurement that is performed for deducing $i$ from $\ket{\psi_i}$. 
Combining Eq. (\ref{H_I:eq}) with inequality (\ref{holevoIneq}) we have 
$H(i|j)\geq H(i) - \chi $ and thus Fano's inequality (\ref{fano1:eq}) yields  
\cite{NCbook} 
\be
H_{\rm bin}(p_{\rm err} )+p_{\rm err} \log_2(N-1)\geq H(i)-\chi.
 \label{fano_holevo:eq}
 %\tag{F2}
\ee
Inequality (\ref{fano_holevo:eq}) suggests that it gets harder to determine ${ i} $ (i.e., the error 
probability increases) as $\chi$ decreases because the accessible information 
also decreases. For the sake of completeness, in  appendix \ref{appendixA} we 
give a few  weaker forms of Fano's inequality.

In closing this section, it is worth keeping in mind that any variant of  Fano's inequality does not directly provide the minimum-error probability, but it gives a bound that relates the error probability to the uncertainty (or equivalently to the accessible information). Fano's inequality provides a general lower bound on error probability which, as we will see in the following sections, may or may not be tight. On the other hand, the actual minimum-error probability will always be the smallest achievable probability of error, which requires using a measurement that has been optimized in the context of the problem under consideration.

\subsection{Minimum-error discrimination} 
\label{sec2b}

Most elementary  courses on quantum theory cover only projective (von Neumann) measurements because most physical systems allow only coarse measurements, i.e., measurements that do not capture fine or detailed information 
about the quantum system under investigation. In contrast, quantum information processing demands precise control, and we are usually interested in the best possible measurement that can be performed. 
In the context of quantum state discrimination we  are interested in the measurement that minimizes the error probability for a given set of states.  
Hence, it is necessary to introduce a comprehensive formalism of quantum measurements.

In quantum physics, an observable is a physical quantity that, at least in principle, can be measured, and it is represented mathematically by a  Hermitian operator on a Hilbert space. To formulate quantum measurements, we consider a set $\{\hat{\Pi}_j\}$ of positive semi-definite (and thus Hermitian) operators that satisfy the completeness relation $\sum_j \hat{\Pi}_j = \hat{\mathbb I}$. 
The operator $\hat{\Pi}_j$ is associated with the outcome $j$, in the sense that it gives the conditional probability to obtain outcome $j$ given  
the measured state $ \hat{\rho}_i$, as follows  
\bea 
P(j|i):= {\rm Tr}(\hat{\Pi}_j \hat{\rho}_i).
\label{Prob_ji}
\eea   
Hence, in the context of quantum physics and information processing, $\{\hat{\Pi}_j\}$ are also referred to as probability operators. 

The conditional probability (\ref{Prob_ji}) is precisely the conditional probability that appears in Eqs. (\ref{cond_ent:def2}) and (\ref{q_y:eq}). 
Throughout this paper we consider pure states,  and $\hat{\rho}_i$ is defined by Eq. (\ref{holevoEq1b}). Hence, we can also write $P(j|i) = \bra{\psi_i} \hat{\Pi}_j\ket{\psi_i}$.
The set of operators $\{\hat{\Pi}_j\}$ forms a probability operator measure (POM), also
known as a positive operator-valued measure (POVM), while we refer to the individual operators in the set as elements of the POM. 

It is worth recalling here that in a von Neumann measurement the possible outcomes are the (real) eigenvalues of the 
observable being measured. One deals with projectors onto the different mutually orthogonal eigenspaces that span the entire Hilbert space.  
As a result, the number of distinct orthogonal projectors (and outcomes) has to be equal to the dimension of the Hilbert space of the quantum system being measured. 
In contrast to projective measurements, there is no such restriction on the number of elements in a POM; 
it can be smaller or larger than the dimension of the state space and may also differ from the number of available preparations of the quantum system’s state.

Any POM corresponds to a realizable measurement that involves extending the Hilbert space where the operators 
$\{\hat{\Pi}_j\}$ are defined to a larger space using an ancillary system, along with the ability to perform unitary transformations and projective measurements \cite{NCbook,Barnettbook}.
Moreover, any measurement can be described by a POM  \cite{NCbook,Barnettbook}. This implies that designing a minimum-error discrimination measurement is an optimization problem: one seeks a set of POM elements $\{\hat{\Pi}_j\}$ that minimize the error probability
\begin{subequations}
\label{MinErr:eqs}
\bea
p_{\rm err} := 
\sum_{i\in\Int_N} \sum_{j\neq i} p_i P(j|i),
%=\frac{1}N \sum_i \sum_{y\neq x}  {\rm Tr}(\hat{\Pi}_j \hat{\rho}_i),
\eea
and maximize the probability of success  
\bea
p_{\rm cor} = 1-p_{\rm err} = \sum_{i\in\Int_N}  p_i P(i|i),
\label{pcor:eq}
\eea
\end{subequations}
where $P(i|i) $ is given by Eq. (\ref{Prob_ji}), for $j=i$.
The necessary and sufficient conditions that must be satisfied by the optimal POM for the general case are known \cite{review1,WeirThesis},  but the details of the  optimal measurement are 
only  known  for certain special cases \cite{helstrom76,AndBarGilHun02,review1,Ban97,Bar01,HauWoo94}.

\subsubsection{Two non-orthogonal qubit states: Helstrom bound}
\label{sec2b1}

Perhaps the most well known result pertains to the discrimination between two equally probable non-orthogonal qubit states, say 
\begin{subequations}
\label{qubit_states:def}
\bea
&&\ket{\psi_0} = \cos(\theta) \ket{0} + \sin(\theta) \ket{1},\\
&&\ket{\psi_1} = \sin(\theta) \ket{0} + \cos(\theta) \ket{1},
\eea
\end{subequations}
where $\{\ket{0},\ket{1}\}$ is an orthonormal qubit basis and $0 \leq \theta\leq \pi/4$. 
The overlap between the two states is 
\bea
\Lambda = \olap{\psi_0}{\psi_1} = \sin(2\theta).  
\label{olap:eq}
\eea
The minimum error probability is  given by the so-called Helstrom bound \cite{helstrom76}
\be
p_{\rm err}^{\min} = \frac{1}{2} \left (1-\sqrt{1-4p_0 p_1 |\Lambda|^2} \right ).
\label{helstrom:eq}
%\tag{ME}
\ee
When $p_0=p_1=1/2$, the optimal measurement is a von Neumann measurement 
with observable $p_0\ket{\psi_0}\bra{\psi_0}-p_1\ket{\psi_1}\bra{\psi_1}$. This is a Hermitian operator and, in principle, it can be observed.  
Using definitions (\ref{qubit_states:def}), one readily obtains that the observable can be also written as 
$(1/2) \cos(2\theta)\left (\ket{0}\bra{0} - \ket{1}\bra{1}\right )$. There are two possible real outcomes, given by the eigenvalues 
$\pm \cos(2\theta)/2$, and these are labeled by $j =0$ and $j = 1$. The corresponding projectors are 
$\hat{\Pi}_0=\ket{0}\bra{0}$ and $\hat{\Pi}_1=\ket{1}\bra{1}$, which lead to guesses $\ket{\psi_0}$ and $\ket{\psi_1}$, respectively.  Note that these two projectors span the entire qubit Hilbert space, while $\hat{\Pi}_i\hat{\Pi}_j = \delta_{i,j}\hat{\Pi}_i$, where  $\delta_{i,j}$ is the Kronecker delta. 
The conditional probabilities of interest are 
\begin{subequations}
\label{Pyx2qubit:eqs}
\bea
&&P(j=i|i) = {\rm Tr}(\hat{\Pi}_i \rho_i) = \cos^2(\theta), \\
&&P(j\neq i|i) = {\rm Tr}(\hat{\Pi}_{j\neq i} \rho_i) = \sin^2(\theta).
\eea
\end{subequations}
The error probability is given by 
\bea
 \frac{1}2 P(j=1|i=0) +\frac{1}2 P(j=0|i=1) = \sin^2(\theta), 
 \label{helstrom2:eq}
\eea
which is equal to the Helstrom bound (\ref{helstrom:eq}), for $\Lambda$ given by Eq. (\ref{olap:eq}) and $p_0=p_1=1/2$.

\subsubsection{Symmetric qubit states}
\label{sec2b2}

Another special case with practical interest considers $N$ equally probable pure qubit states that are related to each other by means of a unitary transformation $\hat{\mathscr U}$, as follows 
\bea
\ket{\psi_i} = \hat{\mathscr U} \ket{\psi_{i-1}} = \hat{\mathscr U}^i \ket{\psi_0},
\eea
where $\hat{\mathscr U}^N =\hat {\mathbb I}$ is the two-dimensional unity operator. Such states are commonly referred to as symmetric states. 
In this case the optimal measurement that minimizes the error probability is the square-root-measurement \cite{review1,WeirThesis,Ban97} corresponding to the probability operators
\bea
\hat{\Pi}_j = \frac{1}N\hat{\rho}^{-1/2}~\hat{\rho}_j~ \hat{\rho}^{-1/2}.
\label{pi_y:eq}
\eea
Let us consider the case of rotation 
\bea
\hat{\mathscr U} = \exp({-{\rm i}\theta \hat{\mathscr{Y}}/2}),
\eea
where $\hat{\mathscr{Y}} := {\rm i} (\ket{1}\bra{0} - \ket{0}\bra{1})$ is the Pauli operator and the non-italic letter i denotes the imaginary unit. (Remember that the italicized letter $i$ is a variable and denotes the label of a quantum state.) Setting $\theta = 2\pi i/N$ and $\ket{\psi_0} = \ket{0}$, we have 
\bea
\ket{\psi_i} = \cos\left ( \frac{i\pi}{N} \right )\ket{0} + \sin\left ( \frac{i\pi}{N} \right )\ket{1}.
\eea

One readily obtains that the average density operator
\bea
\hat{\rho} = \frac{\hat{\mathbb I}}2,
\label{av_rho_pure:eq}
\eea
is the maximally mixed state. Hence, the probability operators (\ref{pi_y:eq}) are simplified to  
\bea
\hat{\Pi}_j = \frac{2}N \ket{\psi_j}\bra{\psi_j}, 
\eea
and the associated conditional probability $P(j|i)$ then reads
 \bea
 P(j|i) = \frac{2}N \cos^2\left [ (i-j)\frac{\pi}{N}\right ],
 \label{pyx_pure:eq}
 \eea
 thereby obtaining 
 \bea
 q_j = \sum_i P(j|i)p_i = \frac{1}{N},  
 \label{py_pure:eq}
\eea
for equally probable states.
Given that the square-root-measurement  minimizes the error probability and maximizes the success probability, we obtain from  Eqs. (\ref{pcor:eq}) and (\ref{pyx_pure:eq}), 
\be
p_{\rm err}^{\rm (min)} = 1-p_{\rm cor}^{\rm (max)}  = 1- \frac{2}{N}. 
\label{helstrom_pure:eq}
%\tag{ME2}
\ee

\subsubsection{Symmetric coherent states}
\label{sec2b3}

The symmetric qubit states discussed in the previous subsection are restricted to a two-dimensional Hilbert space. One may also consider a set of symmetric coherent states 
\begin{subequations}
\label{CoheSet:Eqs}
 \begin{eqnarray}
{\mathbb S}_N \equiv \left \{\ket{\psi_i}:=\ket{\sqrt{\mu}e^{{\rm i} i \delta\varphi}}~:~\delta\varphi:=\frac{2\pi}N,~ i \in \Int_N\right \},\nonumber\\
\label{set_coh:eq}
\end{eqnarray} 
for even $N\geq 2$, where $\mu$ is the mean number of photons and the single-mode 
coherent state $\ket{\psi_i}$ is eigenstate of the bosonic annihilation operator $\hat{a}$, i.e., 
$\hat{a}\ket{\psi_i} = \psi_i \ket{\psi_i}$. It can be expanded in terms of the Fock states 
$\{\ket{n}\}$ as follows 
\bea
\ket{\psi_i}:=e^{-\mu/2} \sum_{n=0}^\infty \frac{\psi_i^n}{\sqrt{n!}}\ket{n}.
\label{alpha_fock:exp}
\eea
In phase space, all the coherent states in ${\mathbb S}_N$ have the same mean number of photons $\mu$, but their phases are distributed around the circle at regular intervals of $\delta\varphi$.

The unitary transformation that maps each state onto its successor 
is the elementary phase-shift operator 
\bea
\hat{\mathscr  U} = e^{{\rm i} 2\pi \hat{a}^\dag \hat{a}/N} =e^{{\rm i} \delta\varphi \hat{a}^\dag \hat{a}} ,
%\hat{\Pi} \hat{\cal S} \hat{\Pi},
\eea
where $\hat{a}^\dag$ is the bosonic creation operator. 
Indeed, one can readily show that 
$\hat {\mathscr  U}^\dag \hat{a} \hat {\mathscr  U} = \hat{a} e^{{\rm i}2\pi/N}=\hat{a} e^{{\rm i} \delta\varphi} $, which implies that 
\bea
\hat {\mathscr U}\ket{\psi_{i \ominus 1}} = 
\ket{\psi_{i\ominus 1} e^{{\rm i}\delta\varphi}} = \ket{\psi_{i}} = \hat{\mathscr  U}^{i}\ket{\psi_{0}} , 
\eea 
where $\hat{\mathscr  U}^{N} = \hat{\mathbb I}$, while $\ominus$ denotes subtraction modulo $N$. 
\end{subequations}

The square-root-measurement is also optimal in the case of symmetric coherent states and minimizes the error probability.  The density operators are given by Eqs. (\ref{holevoEq1b}) and 
(\ref{holevoEq1c}) for $\ket{\psi_i}$ given by Eq. (\ref{set_coh:eq}). 
Using the expansion (\ref{alpha_fock:exp}), one can also show that 
\bea
\hat{\rho} &=& e^{-\mu} \sum_{n=0}^{\infty}\sum_{n^\prime=0}^{\infty} 
\frac{\mu^{(n+n^\prime)/2}}{\sqrt{n!n^\prime!}}\ket{n}\bra{n^\prime} 
\delta(|n-n^\prime|=q N)
\nonumber\\ 
&=&  \sum_{n=0}^{\infty}\sum_{n^\prime=0}^{\infty} 
\sqrt{{\mathscr P}(\mu,n){\mathscr P}(\mu,n^\prime)}
\ket{n}\bra{n^\prime} 
\delta(|n-n^\prime|=q N), 
\nonumber\\
\label{av_dm_coh}
\eea
for $q\in \mathbb{N}_0$, where $\delta(\cdot)$ has non-zero contributions only
for $|n-n^\prime|=0,N,2N, \ldots$, and ${\mathscr P}(\mu,n)$ denotes 
the Poisson distribution.
  
Unlike symmetric qubit  states, symmetric coherent states do not allow an easy derivation of a general expression for the minimum-error probability or the Holevo bound $\chi$. Instead, one has to 
resort to numerical calculations. Perhaps the only exceptions are the case of two coherent states, where  the 
minimum error probability is given by the Helstrom bound (\ref{helstrom:eq}), with $\Lambda$ 
the overlap between the two coherent states, as well as the special cases discussed in appendix \ref{appendixB}.

\section{Results and Discussion} 
\label{sec3}

In this section our aim is to compare the error bounds obtained by inequalities (\ref{fano1:eq}) and (\ref{fano_holevo:eq}) to the minimum-error probability for the three cases of subsections (\ref{sec2b1}) -  (\ref{sec2b3}). 
To facilitate this discussion, we will refer to $p_{\rm err}^{(1)}$ and $p_{\rm err}^{(2)}$ as the solutions of inequalities (\ref{fano1:eq}) and (\ref{fano_holevo:eq}), respectively. 
The main problem with inequality (\ref{fano1:eq}) is that its solution depends on the conditional entropy  $H(i|j)$, which in turn depends on the measurement  and the decision-making strategy under consideration. Therefore, the first question  we have to answer is which measurement we should consider in the comparison? 

One can easily check that the left-hand side  of  inequality (\ref{fano1:eq}) is a concave 
function of the error probability and has at most two common points with $H(i|j)$, 
which define the lower and upper bounds for the interval of values of $p_{\rm err}$ 
in which the inequality is satisfied. 
The lower bound is always present  and decreases (approaching $p_{\rm err} = 0$) for decreasing  $H(i|j)$ [or equivalently for increasing $I(i:j)$]
The necessary condition for a POM measurement to minimize  $H(i|j)$ [maximize $I(i:j)$]  is known, but to the best of our knowledge, the sufficient condition has not been found yet \cite{Ban97}. 
It is also known \cite{Ban97}, that for the cases of subsections (\ref{sec2b1}) -  (\ref{sec2b3}), the minimum-error discrimination measurement is  the one that satisfies the 
necessary condition for maximization of $I(i:j)$ [minimization of $H(i|j)$]. In the following discussion, therefore, we will consider the solutions of inequality (\ref{fano1:eq})  
for this measurement only, as it is expected to be closer to the minimum-error probabilities in each case.

\subsection{Two qubit states}
\begin{figure}
\centering\includegraphics[width=8cm]{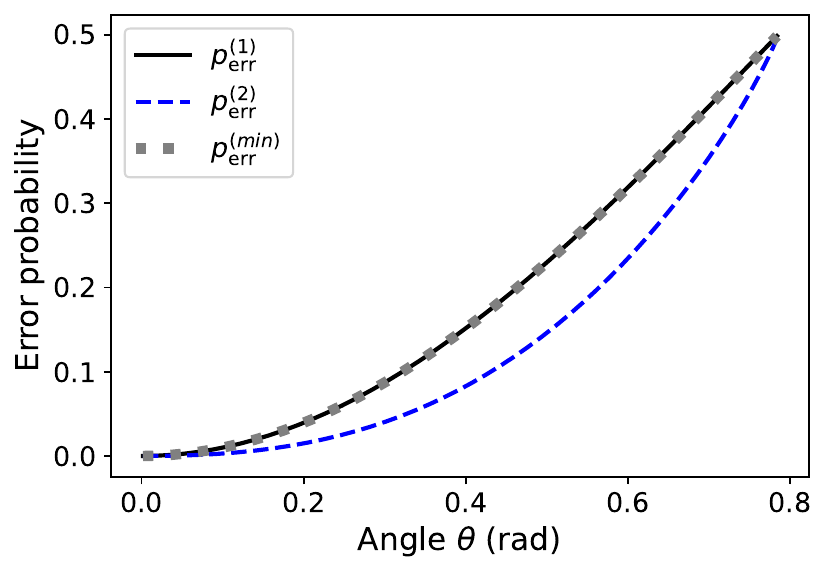}
\caption{Error probabilities for the discrimination of two equally probable non-orthogonal qubit states as a function of the angle $\theta$ between them. The dotted gray line shows the Helstrom bound, as given by Eq. (\ref{helstrom:eq}), versus the  bounds obtained through the solution of Fano's inequality. The black solid line pertains to inequality (\ref{fano1:eq}), with $H(i|j)$ given by Eq. (\ref{Eq:f1b}), and   
the  blue dashed line pertains to inequality (\ref{fano_holevo2:eq}). 
}
\label{figure1}
\end{figure}

Let us begin with the simplest case (which is still of practical interest): the discrimination of $N=2$ equally probable non-orthogonal qubit states.  
For the optimal measurement discussed in Sec. \ref{sec2b1}, and using Eqs. (\ref{Pyx2qubit:eqs}), (\ref{cond_ent:def2}), and (\ref{q_y:eq}), we obtain  
\be
H(i|j) = - 
 \cos^2(\theta)\log_2[ \cos^2(\theta)]   - \sin^2(\theta)\log_2[\sin^2(\theta)]. \label{Eq:f1b}
 % \tag{F1B}
\ee
In view of Eqs. (\ref{helstrom:eq}) and (\ref{helstrom2:eq}), Eq. (\ref{Eq:f1b}) is equal to $H_{\rm bin}(p_{\rm err}^{\rm (min)})$, thereby obtaining for Fano's inequality (\ref{fano1:eq}) 
\be
H_{\rm bin}(p_{\rm err}) \geq H_{\rm bin}(p_{\rm err}^{\rm (min)}), \label{Eq:f1c}
 % \tag{F1B}
\ee
which implies that  $p_{\rm err}^{(1)} = p_{\rm err}^{\rm (min)}$ for all values of $\theta$. That is, in the case of two equally probable non-orthogonal states, the solution of Fano's inequality  (\ref{fano1:eq}) yields the minimum error probability when the conditional entropy $H(i|j)$ is calculated for the minimum-error discrimination measurement discussed in Sec. \ref{sec2b1}. 
This is also confirmed numerically, as depicted in Fig. \ref{figure1}. 
For $\theta = 0$, the states under consideration become orthogonal [see Eqs. (\ref{qubit_states:def})] and the error probability is zero because perfect discrimination is possible in this case. On the other hand, for $\theta=\pi/4$ the two states are identical and the best one can do is random guessing, with error probability equal to $1/2$. 

The variant of Fano's inequality (\ref{fano_holevo:eq}) reads for the case  under consideration 
\begin{subequations}
\label{fano_holevo2:eq}
\be
H(p_{\rm err} )\geq 1+\lambda_+\log_2(\lambda_+)+\lambda_-\log_2(\lambda_-),
%\tag{F2b}
\ee
where 
\bea
\lambda_\pm = \frac{1}2 \left ( 1\pm \Lambda \right ), 
\eea
\end{subequations}
are the eigenvalues of the density matrix
\bea
\hat{\rho} = \frac{1}2 \left ( \ket{\psi_0}\bra{\psi_0}+\ket{\psi_1}\bra{\psi_1}\right ).
\eea
As shown in Fig. \ref{figure1}, inequality (\ref{fano_holevo2:eq}) yields an error probability $p_{\rm err}^{\rm (2)}<p_{\rm err}^{\rm (min)}$. 

\subsection{Symmetric qubit states}

Let us turn now to the case of $N\geq 3$ equally probable symmetric qubit states. Using Eq.  (\ref{py_pure:eq}),  and taking into account the fact that Eq. (\ref{pyx_pure:eq}) depends only on the difference $d:=|i-j|$ and is symmetric with respect to $i$ and $j$,
we arrive at the following expression for the conditional entropy on the r.h.s. of inequality (\ref{fano1:eq}) 
\bea
H(i|j)&:=& - \frac{2}{N} \sum_{d = 0}^{N-1} \cos^2\left ( \frac{d\pi}{N}\right )  \log_2\left [\frac{2}{N} \cos^2\left ( \frac{d\pi}{N}\right ) \right ].\nonumber \\
\label{Hxy:Nqubit:eq}
\eea
We observe that, for  $N=3$ qubit states, $H(i|j) = p_{\rm err}^{\rm min}+H_{\rm bin}(p_{\rm err}^{\rm min})$ and 
Fano inequality (\ref{fano1:eq}) simplifies to 
\bea
H_{\rm bin}(p_{\rm err}) +p_{\rm err} \geq H_{\rm bin}(p_{\rm err}^{\rm (min)})+p_{\rm err}^{\rm (min)},
\eea
where $p_{\rm err}^{\rm (min)} = 1/3$ as  given by Eq. (\ref{helstrom_pure:eq}). 
The expression on the left-hand side of the inequality is a monotonically increasing function of $p_{\rm err}$,  and the lowest error probability is $p_{\rm err}^{(1)} = p_{\rm err}^{\rm (min)}$, 
which is in agreement with our simulations.  
This is one more example, besides $N=2$, where the Fano's inequality yields an error-probability bound that agrees with the minimum-error probability. The natural question that arises, therefore, is whether 
the same happens for $N>3$ symmetric qubit states.

Before answering this question let us also consider the variant of Fano's inequality 
(\ref{fano_holevo:eq}), which, using Eqs. (\ref{holevoEq1a}), (\ref{av_rho_pure:eq}), becomes 
\be
H(p_{\rm err} )+p_{\rm err} \log_2(N-1)\geq \log_2(N)-1.
\label{fano_holevo_pure:eq}
%\tag{F2q}
\ee

\begin{figure}
\centering\includegraphics[width=8cm]{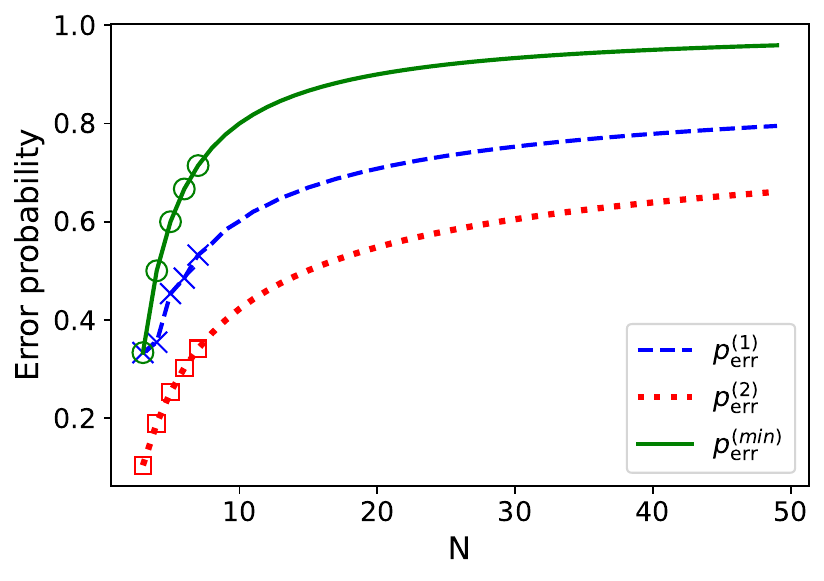}
\caption{Comparison of bounds on error probability for $N>2$ symmetric qubit states. The green solid line shows the minimum-error discrimination probability as given by Eqs. (\ref{helstrom_pure:eq}). 
The blue dashed  line shows the  bound obtained through the 
solution of Fano inequality (\ref{fano1:eq}) with $H(i|j)$ given by 
(\ref{Hxy:Nqubit:eq}), and the red dotted line shows the solution of Fano inequality (\ref{fano_holevo_pure:eq}). The error probabilities are also shown with symbols for the first five values of $N>2$.  
}
\label{figure2}
\end{figure}

The solution of inequality (\ref{fano1:eq}), together with the solution of inequality (\ref{fano_holevo_pure:eq}) and the minimum-error probability (\ref{helstrom_pure:eq}), 
are plotted in Fig. \ref{figure2} for $N\geq 3$. 
Clearly, the differences between the lower error-probability bounds obtained by  
Fano's inequality and the minimum-error probability increase with the number of qubit states for $N>3$, while the predictions of Fano's inequalities are always well below  the minimum-error probability.

\subsection{Symmetric coherent states} 

\begin{figure}
\centering{\includegraphics[width=8cm]{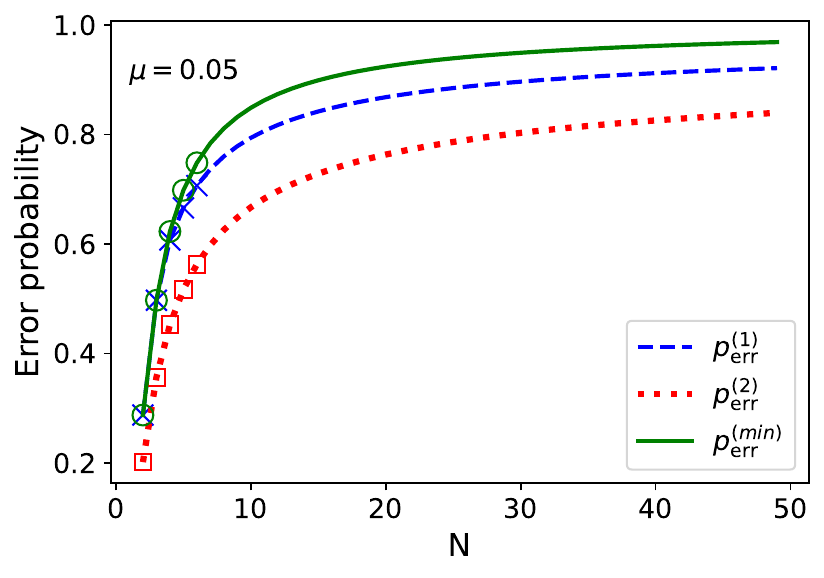}
\includegraphics[width=8cm]{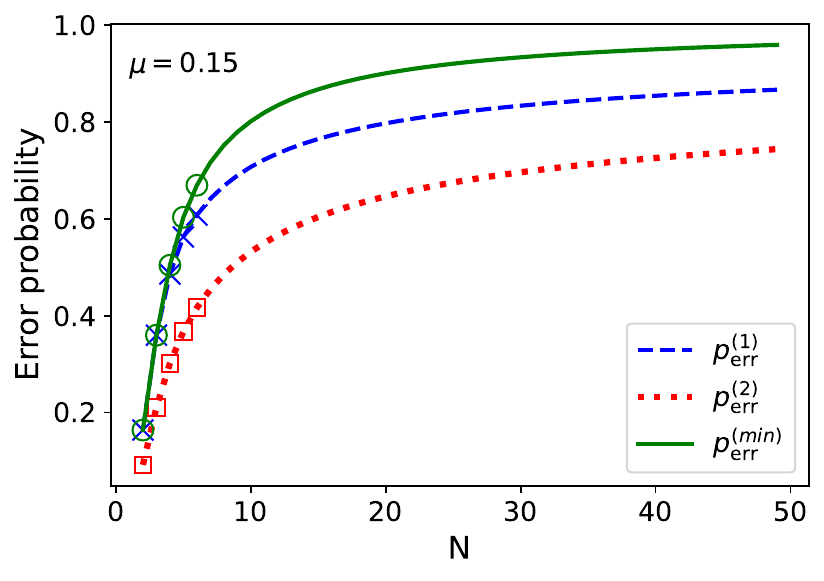}}
\caption{Comparison of bounds on error probability for $N\geq 2$ symmetric coherent states. The green  line shows the minimum-error probability obtained from Eqs. (\ref{MinErr:eqs}) and (\ref{pi_y:eq}). 
The blue dashed line shows the  bound obtained through the 
solution of Fano's inequality (\ref{fano1:eq}) with $H(i|j)$ calculated for the square-root measurement, and the red dotted line shows the solution of Fano's inequality (\ref{fano_holevo:eq}). The error probabilities are also shown with symbols for the first five values of $N\geq 2$.  
}
\label{figure3}
\end{figure}

Having discussed the case of symmetric qubit states, we turn now to discuss the case of 
$N\geq 2$ symmetric coherent states. 
In Fig. \ref{figure3} we plot  the minimum-error probability, together 
with the solutions of the two variants of Fano's inequality discussed above, for two different values of the mean number of photons. 
In analogy to the findings for the qubit states, we see that for $N=2$ and $N=3$,  the minimum-error probability is very close (but not identical) to the error bound of Fano's inequality (\ref{fano1:eq}) when $H(i|j)$ is calculated for the square-root measurement. Larger  deviations between the two  error probabilities appear for $N> 4$, and they seem to increase as the mean number of photons, $\mu$, increases. 
The error bounds obtained by Fano's inequality (\ref{fano_holevo:eq}) are always 
well below the minimum-error probability. 

\subsection{Comparison of different cases}

\begin{figure}
\centering{\includegraphics[width=8.5cm]{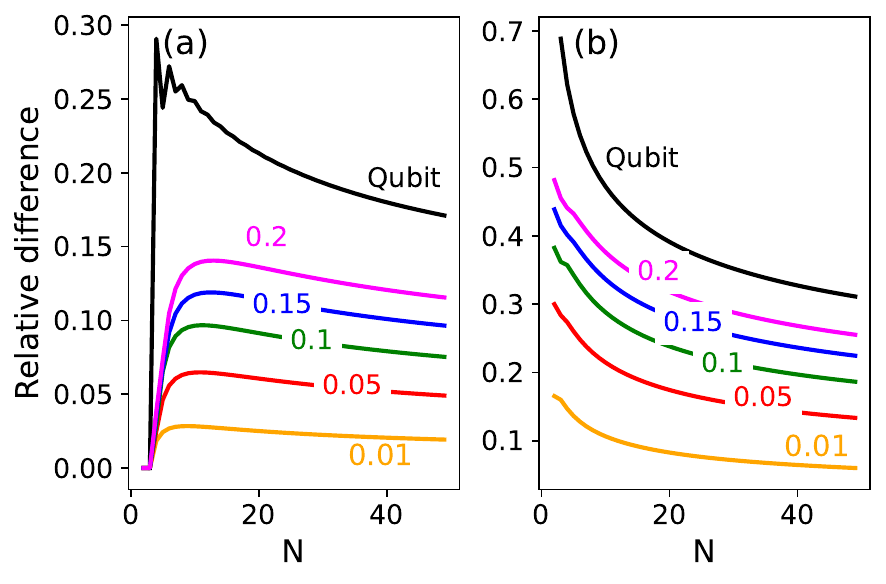}}
\caption{The relative difference between the minimum error probability and the 
solutions of (a) Fano's inequality (\ref{fano1:eq}) and (b) Fano's inequality (\ref{fano_holevo:eq}) for symmetric qubit states and symmetric coherent states with various mean number of photons (shown next to the curves).  
In each case,  the entropy $H(i|j)$  in the left-hand side of inequality (\ref{fano1:eq}) has been obtained for the minimum-error discrimination measurement. 
}
\label{figure4}
\end{figure}

To facilitate the comparison between the different cases, we introduce the relative difference 
\bea
{\cal D}_l = \frac{|p_{\rm err}^{(l)}-p_{\rm err}^{(\rm min)}|}{p_{\rm err}^{(\rm min)}},
\eea
for $l=1,2$.
In Fig. \ref{figure4} we plot ${\cal D}_l $ as a function of $N$ for symmetric qubit states as well as for symmetric coherent states of different values of $\mu$. 
First of all, for weak coherent states $(\mu\leq 0.2)$, the observed differences for 
symmetric qubit states are well above the ones for coherent states. The differences decrease 
with decreasing $\mu$, while for fixed $\mu$ the differences decrease with increasing 
$N$, as the error probability approaches 1 in all of the cases. 
In general, ${\cal D}_2>{\cal D}_1$, which was to be expected 
because inequality (\ref{fano_holevo:eq}) is a weaker variant of Fano's inequality (\ref{fano1:eq}). However, this does not diminish the usefulness of inequality (\ref{fano_holevo:eq}), which rests on its generality, in the sense that it does not depend on the applied measurement.
In many cases of practical interest one  has to make statements about the efficiency or the security of a quantum-information-processing scheme without making assumptions about the measurements involved in the process.  
It is precisely in such scenarios where  inequality (\ref{fano_holevo:eq}) can be particularly useful.

In all of the scenarios we have discussed for qubit and coherent states, the error bound obtained through Fano's inequality is below the minimum-error probability. This may seem weird, 
but as mentioned above, Fano's inequality is not a direct calculation of the error probability, but rather it provides a  bound on the conditional entropy, which in turn relates to error probability. Its usefulness does not rest on providing an exact value for the error probability, but rather a guaranteed  bound, which in some cases can be tight (e.g., see the  cases of two and three qubit or coherent states discussed above). 
On the other hand, the minimum-error probability refers to the smallest possible probability of error achievable by the optimal measurement, where the decision rule has been optimized. 
Fano's inequality (\ref{fano_holevo:eq})  does not depend at all on the applied measurement, whereas Fano's inequality (\ref{fano1:eq}) depends on the applied measurement only through
 $H(i|j)$, on the right-hand side.

\section{Conclusions}

We have discussed in detail the minimum-error discrimination and Fano's inequality for special cases 
of practical interest, involving  two or more equally probable non-orthogonal 
quantum states. For $N=2$ and 
$N=3$ qubit states, we have obtained analytic results that show that the lower error-probability bounds obtained through Fano's inequality are identical to the minimum-error probability. In the case of $N=2$ and 
$N=3$ coherent states, the lower error-probability bounds  are very close, but not identical, to the minimum-error probability. For $N>3$  symmetric qubit or coherent states, all variants of 
Fano's inequality yield results that are well below the minimum-error probability. 

The present analysis sheds some light on possible connections of Fano's inequality to 
minimum-error discrimination and may serve as a guide for related lectures  in the framework of undergraduate and graduate courses on quantum physics and information processing. Moreover, the examples that are amenable to analytic solutions may be  useful for educators to demonstrate fundamental aspects of quantum information theory  (such as the theoretical description of quantum measurements and the calculation of entropy and  information gain) and point out the main differences between qubit and coherent states so that the students get a better understanding of the concepts. 

For educational purposes, in all the scenarios we considered, the quantum states were assumed to be equally probable, which allowed us to obtain analytical solutions in certain cases. Our work can also be extended to the case of unequally probable states, although deriving analytical results will not be possible in most cases, and one has to resort to numerical solutions. Moreover, the optimal measurements that minimize the error probability may differ than the ones discussed here. This extension is left as an exercise for students interested in exploring quantum state discrimination more deeply. 

\appendix % Add  * if there's only one appendix.

\section{Other variants of Fano's inequality}
\label{appendixA}

 Using the fact that the binary entropy always satisfies  $H_{\rm bin} (p_{\rm err})\leq 1$, 
 one readily obtains from  inequalities (\ref{fano1:eq}) and (\ref{fano_holevo:eq})
 \bea
 p_{\rm err}\geq \frac{H(i|j)-1}{\log_2(N-1)},
 \eea
 and 
  \bea
 p_{\rm err}\geq \frac{H(i)-\chi}{\log_2(N-1)},
 \eea
 respectively. These are weaker variants of Fano inequalities (\ref{fano1:eq}) and (\ref{fano_holevo:eq}), and they are expected to give a bound that is well below the ones discussed in the main sections of this work. 
 
\section{Special cases of symmetric coherent states}
\label{appendixB}

There is a case where the  density matrix (\ref{av_dm_coh}) is simplified considerably. The Poisson distribution is discrete with non-negligible probabilities mainly for values of $n$ 
in some interval $[n_{\rm min} , n_{\rm max}]$. 
If $N\geq n_{\rm max} - n_{\rm min}$, then  
the function $\delta(\cdot)$ essentially reduces to Kronecker delta $\delta_{n,n^\prime}$, and the density operator becomes diagonal for all practical purposes: 
\bea
\hat{\rho} \simeq e^{-\mu}  \sum_{n=0}^{\infty} 
\frac{\mu^n}{n!}\ket{n}\bra{n} :=
\sum_{n=0}^{\infty} {\mathscr P}(\mu,n)\ket{n}\bra{n}, 
\label{rho_diag:eq}
\eea  
which is independent of the phase slices.  
Hence,  we can obtain analytic expressions for the probability operators for the square-root measurement and the corresponding minimum error probability. 
Using Eqs.  (\ref{pi_y:eq})  and (\ref{rho_diag:eq}) we have 
\bea
\hat{\Pi}_j = \frac{e^{-\mu}}N \sum_{n,m} \frac{[{\mathscr P}(\mu,n){\mathscr P}(\mu,m)]^{-1/2} \psi_j^{n} (\psi_j^{*})^m}{\sqrt{n!m!}} \ket{n}\bra{m},\nonumber\\
\eea
and thus we obtain for the conditional probability 
\bea
P(i|i) &=& {\rm Tr} (\hat{\Pi}_i \hat{\rho}_i) = \frac{1}N. 
\eea
 Equation  (\ref{pcor:eq}) yields 
\bea
p_{\rm err}^{\rm (min)} = 1-p_{\rm cor}^{\rm (max)}  = 1- \sum_{x\in\Int_N}  p_i P(i|i) = 1- \frac{1}{N}.
\label{helstrome_coh_diag:eq}
\eea  
Moreover,  using  (\ref{rho_diag:eq}) in Eq. (\ref{holevoEq1a}), we have 
\bea
\chi = \sum_n {\mathscr P}(\mu,n) \log_2 [{\mathscr P}(\mu,n)],
\label{fano_holevo_coh:eq}
\eea
which is the entropy of the Poisson distribution, and  
in the limit of $\mu\gtrsim 10$, it is well approximated by $H_{\infty}(\mu) = \log_2(\sqrt{2\pi e \mu})$.

\begin{acknowledgments}
This research was co-funded by the European Union under the Digital Europe Program grant agreement number 101091504.
\end{acknowledgments}

\section*{AUTHOR DECLARATIONS}

\subsection*{Conflict of Interest}
The author has no conflicts to disclose.

%\subsection*{Data Availability Statement}
%The data that support the findings of this study are available within the article [and its supplementary material].


\begin{thebibliography}{99}
\bibitem{footnote}
Note that the binary entropy provides the average number of bits needed for the particular task. 
One needs 1 bit on average only in the case of balanced distribution with $p_{\rm err}=1/2$. When $p_{\rm err}\neq 1/2$, however, and considering a sequence of many messages, one can design a code that uses fewer bits on average, by assigning shorter codewords to more probable messages and larger codewords to the less probable ones. 
This is related to  Shannon’s first noiseless coding theorem and for more information  the interested reader may refer, for instance, to Chapter 12 of \cite{NCbook}, Chapter 2 of \cite{Wildebook}, and the references therein.
\end{thebibliography}
\end{document}